\documentclass[twocolumn]{aastex631}

\begin{document}

\title{An independent determination of the distance to supernova SN\,1987A by means of the light echo AT\,2019xis}

\author[0000-0001-7101-9831]{Aleksandar Cikota}
\affiliation{Gemini Observatory / NSF's NOIRLab, Casilla 603, La Serena, Chile; aleksandar.cikota@noirlab.edu}

\author[0000-0003-4928-6698]{Jiachen Ding}
\affiliation{Department of Atmospheric Sciences, Texas A\&M University, College Station, TX 77843, USA}

\author[0000-0001-7092-9374]{Lifan Wang}
\affiliation{George P. and Cynthia Woods Mitchell Institute for Fundamental Physics and Astronomy, Texas A\&M University, 4242 TAMU, College Station, TX 77843, USA}

\author[0000-0003-1637-9679]{Dietrich Baade}
\affiliation{European Organisation for Astronomical Research in the Southern Hemisphere (ESO), Karl-Schwarzschild-Str.\ 2, 85748 Garching b.\ M\"{u}nchen, Germany}

\author[0000-0002-7671-2317]{Stefan Cikota}
\affiliation{Centro Astronónomico Hispano en Andalucía, Observatorio de Calar Alto, Sierra de los Filabres, 04550 Gérgal, Spain}

\author[0000-0002-4338-6586]{Peter H\"oflich}
\affiliation{Department of Physics, Florida State University, Tallahassee, FL 32306-4350, USA}

\author[0000-0003-0733-7215]{Justyn Maund}
\affiliation{Department of Physics and Astronomy, The University of Sheffield, Hicks Building, Hounsfield Road, Sheffield, S3 7RH, UK}

\author{Ping Yang}
\affiliation{Department of Atmospheric Sciences, Texas A\&M University, College Station, TX 77843, USA}



\begin{abstract}
Accurate distance determination to astrophysical objects is essential for the understanding of their intrinsic brightness and size. The distance to SN\,1987A has been previously measured by the expanding photosphere method, and by using the angular size of the circumstellar rings with absolute sizes derived from light curves of narrow UV emission lines, with reported distances ranging from 46.77 kpc to 55 kpc. In this study, we independently determined the distance to SN\,1987A using photometry and imaging polarimetry observations of AT\,2019xis, a light echo of SN\,1987A, by adopting a radiative transfer model of the light echo developed in \citet{Ding2021}. We obtained distances to SN\,1987A in the range from 49.09 $\pm$ 2.16 kpc to 59.39 $\pm$ 3.27 kpc, depending on the interstellar polarization and extinction corrections,  which are consistent with the literature values. This study demonstrates the potential of using light echoes as a tool for distance determination to astrophysical objects in the Milky Way, up to kiloparsec level scales.
\end{abstract}



\section{Introduction}

Distance determination to astrophysical objects is fundamental for improving our knowledge about these objects. Without knowing the distance to objects, their intrinsic brightness and size cannot be determined. Furthermore, accurate distances to extragalactic objects are crucial for the determination of the Hubble constant, H$_0$, and measuring the expansion rate of the Universe \citep{1998AJ....116.1009R, 1999ApJ...517..565P}. The distance to the Large Magellanic Cloud (LMC) is important because the Cepheids in the LMC are used as an anchor for the cosmic distance ladder \citep[e.g.][]{2019ApJ...876...85R}.
Probably the most accurate distance to the LMC within 1 per cent was reported by \citet{2019Natur.567..200P}, who determined the distance, d = 49.59 $\pm$ 0.09 (statistical) $\pm$ 0.54 (systematic) kpc. They applied the interferometry-based surface brightness color relation derived from single late-type stars to fully detached eclipsing binaries with similar spectral types. 

SN\,1987A, which exploded in the LMC, has also a relatively well determined distance.
\citet{1987ApJ...320L..23B} determined the distance to SN\,1987A of 55 $\pm$ 5 kpc by matching a distance-independent photometric angular radius of the photosphere to a distance-dependent spectroscopic angular radius, which is known as the Baade method \citep{1926AN....228..359B} or the expanding photosphere method \citep{1974ApJ...193...27K}.
Using the same method, \citet{1988PASA....7..434H} obtained a distance of 48 $\pm$ 2 kpc based on full non-LTE models by fitting a spectral sequence of 8 spectra obtained during the first 7 months. 
\citet{1989ApJ...347..771E} developed a model atmosphere for SN\,1987A and determined the distance to SN\,1987A of 49 $\pm$ 6 kpc using the expanding photosphere method.
\citet{1991A&A...249...36H} also applied the Baade's method for distance determination and derived a distance of 52.3 $\pm$ 1.5 kpc.
\citet{1991ApJ...380L..23P} have compared the angular size of the circumstellar rings of SN\,1987A in the Hubble Space Telescope (HST) images with absolute sizes determined from light curves of narrow UV emission lines and derived a distance of 51.2 $\pm$ 3.1 kpc to SN\,1987A. They also determined a distance of 50.1 $\pm$ 3.1 kpc to the center of the LMC using radial velocities of the SN relative to the center of LMC.
\citet{1992ApJ...395..366S} used IR photometry of SN\,1987A and determined a distance of 49 $\pm$ 3 kpc using the expanding photosphere method.
\citet{1994ApJ...425...51G} revisited the supernova-ring method and the measurements by \citet{1991ApJ...380L..23P} and determined a distance to the SN of 52.7 $\pm$ 2.6 kpc and to the LMC of 53.2 $\pm$ 2.6 kpc.
Based on new measurements of the ionized-emission light curves and the angular size of the ring around SN\,1987A, \citet{1995ApJ...452..189G} derived an upper limit for the distance to the SN of D $<$ 46.77 $\pm$ 0.76 kpc.
\citet{1995ApJ...438..724C} investigated the light echoes from the inner region of SN\,1987A to define the 3D geometry of the circumstellar nebula and used the \citet{1991ApJ...380L..23P} method to determine the distances to the SN and LMC of d$_{\rm SN}$ = 51.7 $\pm$ 3.1 kpc and d$_{\rm LMC}$=51.9 $\pm$ 3.1 kpc, respectively.
\citet{2002ApJ...574..293M} conducted a detailed spectroscopic analysis of SN\,1987A and modeled the observed spectra from day 1 to 81 using hydrodynamical models. Furthermore, they used the spectral-fitting expanding atmosphere method to derive the distance to SN\,1987A of 50 $\pm$ 5 kpc.

Distances to variable sources can also be determined by means of light echoes \citep[e.g.][]{1973A&A....25..445T,1994ApJ...433...19S,1995A&A...293..889P,2000A&A...357L..25P,2016ApJ...825...15H}. 

When light from a bright source, such as a supernova, reaches a cloud of dust, some of the light is scattered in different directions and eventually reaches an observer on Earth. 
\citet{1994ApJ...433...19S} proposed to measure distances to historic supernovae by observing highly polarized circles of scattered light around these objects. Because the maximum polarization occurs for a scattering angle of 90 degress, the radius of the light echoes can be used as a ruler and corresponds to the length $ct$, where $t$ is the time since the supernova exploded and $c$ is the speed of light.
Using that technique, \citet{2008AJ....135..605S} found the geometric distance to V838 Monocerotis, an unusual variable star with a light echo that appeared after an outburst in 2002. 
Even if the scattering angle of maximum polarization is not 90 degrees, the method may still be used if the scattering angle is known. 
\citet{2014A&A...572A...7K} obtained imaging polarimetry of the reflection nebula around the long-period Galactic Cepheid RS Puppis and utilized two polarization models (one based on Milky Way dust mixture and the other assuming Rayleigh scattering) to retrieve  the scattering angle from the degree of polarization. By taking into account the dust distribution in the nebula, they adjusted a model of the phase lag of photometric variations over specific nebular features to calculate the distance to RS Puppis.

We determined the distance to SN\,1987A by means of a light echo of SN\,1987A, AT\,2019xis. 
We obtained imaging polarimetry observations of AT\,2019xis, and in combination with the publicly available photometry of AT\,2019xis applied the radiative transfer model of the SN 1987A light echo developed in \citet{Ding2021} to determine the distance to SN\,1987A.

\begin{figure*}
\centering
\includegraphics[width=\textwidth]{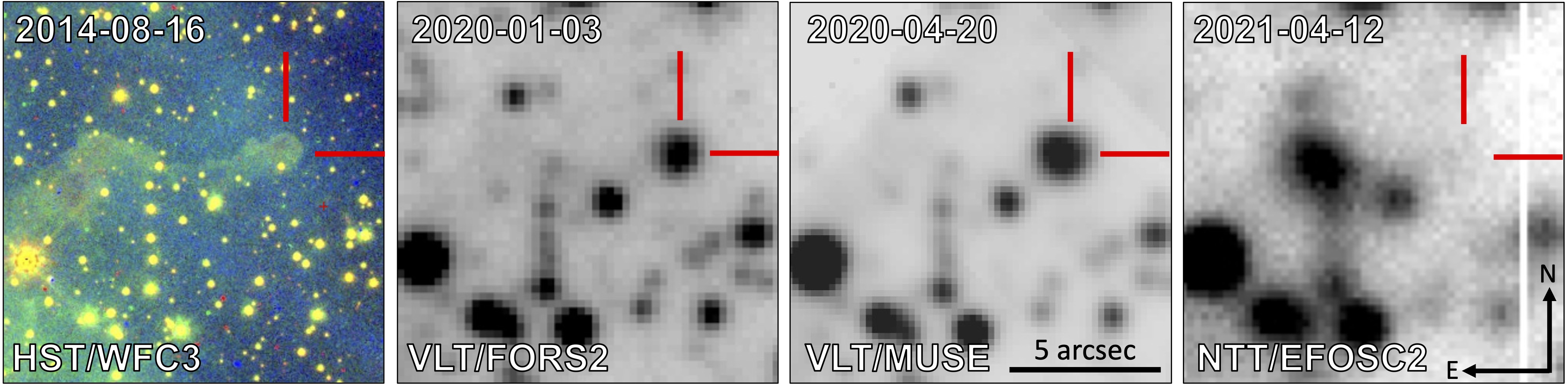}
\caption{Light echo AT\,2019xis observed on January 3rd, 2020 with FORS2 in the $V$ band (this work), April 20th, 2020 with the Multi Unit Spectroscopic Explorer (MUSE, Prog. ID 2104.D-5041(A), PI Kuncarayakti) and April 12th, 2021 with the ESO Faint Object Spectrograph and Camera (EFOSC2) in the $V$ band (as part of PESSTO, \citealt{2015A&A...579A..40S}), compared to a Hubble Space Telescope (HST) archival image (Prog. ID. 13401, PI Fransson) observed in the $F814W$, $F606W$ and $F502N$ bands with the Wide Field Camera 3 (WFC3). The red lines mark the position of AT\,2019xis. The light echo propagated eastward from SN\,1987 which is located 4.05' west from AT\,2019xis.}
\label{fig:lightechoimage}
\end{figure*}

The light echo, AT\,2019xis ($\alpha$ = 05:36:13.700, $\delta$ = -69:16:24.70), was discovered by the Optical Gravitational Lensing Experiment (OGLE) on 2019-10-15 \citep{2019TNSTR2673....1G}. It was initially classified on 2019-12-28 by ePESSTO+ as a SN\,1987A-like Type II supernova \citep{2019TNSAN.160....1A} at $\sim$ 70 days after the shock breakout. 
However, AT\,2019xis was later re-classified as a light echo of SN\,1987A  from the tip of a dust cloud \citep{2019TNSAN.163....1T}, which is also visible in archival HST images, located at an angular distance of 4.05' from SN\,1987A. Fig.~\ref{fig:lightechoimage} shows images of AT\,2019xis observed at three different epochs compared to a HST image.

In Sect.~\ref{sect:observations} we describe the observations and data reduction, in Sect.~\ref{sect:model} the model and method for the distance determination, and in Sect.~\ref{sect:results} we report the results and discuss the present findings in comparison with the literature.

\section{Data and observations}
\label{sect:observations}

\subsection{Imaging polarimetry}

We obtained imaging linear polarimetry of the light echo AT\,2019xis with the FOcal Reducer/low dispersion Spectrograph 2 (\citealt{1998Msngr..94....1A}, FORS2) mounted on the primary focus of European Southern Observatory's (ESO) Very Large Telescope (VLT) Antu (Prog. ID 2104.C-5031(A), PI Cikota). The data was taken on January 3rd, 2020, during good weather conditions. 
The object was at an airmass between 1.4 to 1.5 during the observations. The DIMM seeing was between $\sim$ 0.8-1.0", and the Moon was set.
 
The observations were taken in three different passbands, the FORS2 V\_HIGH ($\lambda_0$ = 555 nm, FWHM = 123.2 nm), R\_SPECIAL ($\lambda_0$ = 655 nm, FWHM = 165 nm) and I\_BESS ($\lambda_0$ =  768 nm, FWHM = 138 nm) filters, at four half-wave plate (HWP) angles of 0, 22.5, 45, and 67.5 degrees. The redundancy of four HWP angles reduces the flat-fielding issue and cancels out other instrumental effects (see \citealt{Patat2006}). The sequence was repeated twice in the $V$ and $I$ bands to increase the signal-to-noise ratio, and taken only once in the $R$ band. 

The brightness of 2019xis at the time of the polarimetry observations was 18.93 $\pm$ 0.03 mag in the $I$ band, as measured by the Optical Gravitational Lensing Experiment (OGLE) project \citep{2015AcA....65....1U}. The exposure times used were 400 sec, 250 sec and 200 sec, which led to a signal-to-noise ratio of $\sim$80, $\sim$85 and $\sim$100 per exposure in the $V$, $R$ and $I$ band, respectively.

The linear polarization was calculated following the standard approach \citep{FORS2manual, 2017MNRAS.464.4146C}.  
We measured the flux of the light echo in the ordinary and extraordinary beams at all HWP angles using the aperture photometry function from PythonPhot\footnote{\href{https://github.com/djones1040/PythonPhot}{https://github.com/djones1040/PythonPhot}}. We used an aperture radius of 1.3", with the inner and outer sky radii of 1.5" and 2.5", respectively. Tests with slight variations of the aperture and sky radii produced consistent results within the 1$\sigma$ uncertainties.

The normalized Stokes $q$ and $u$ parameters were calculated as described in the FORS2 User Manual (\citealt{FORS2manual}, see also \citealt{2017MNRAS.464.4146C}, and \citealt{2022MNRAS.509.6028C}):
\begin{equation}
\label{eq:stokes}
\begin{array}{l}
q = \frac{2}{N} \sum_{i=0}^{N-1} F (\theta_i) \cos (4\theta_i), \\ \\
u = \frac{2}{N} \sum_{i=0}^{N-1} F (\theta_i) \sin (4\theta_i),
\end{array}
\end{equation}
where $N$ is the number of the half-wave retarder plate angles, $\theta_i = 22.5 ^\circ \times i$, with $0 \leq i \leq 3$, and $F(\theta_i)$ is the normalized flux difference between the ordinary ($f^o$) and extraordinary ($f^e$) beams:

\begin{equation}
\label{eq:normfluxdiff}
F (\theta_i) = \frac{f^o (\theta_i) - f^e (\theta_i)}{f^o (\theta_i) + f^e (\theta_i)}.
\end{equation}

The polarization position angles of the raw measurements have been corrected for the half-wave plate zero angle chromatic dependence \citep[Table 4.7 in][]{FORS2manual}.

Finally, we calculated the degree of linear polarization, $p$, and the polarization angle, $\theta$:
\begin{equation}
p=\sqrt{q^2+u^2} , \hspace{10pt}
\theta = \frac{1}{2} arctan(u/q) ,   
\end{equation}
and apply a polarization bias correction following \citet{1997ApJ...476L..27W}. The imaging polarimetry results are summarized in Table~\ref{tab:impol_results}, and shown in Figure~\ref{fig:VRIMag1}.

\begin{table*}
	\centering
	\small
	\caption{\label{tab:impol_results} The imaging polarimetry measurements of the light echo AT\,2019xis. The columns labeled $q_{\rm ISP}$ and $u_{\rm ISP}$ denote the Stokes $q$ and $u$ parameters of the interstellar polarization, as determined from field stars. The column labeled $p_{\rm ISPcorr}$ denotes the interstellar polarization corrected polarization of AT,2019xis. Both the observed polarization and the ISP corrected polarization have been corrected for the polarization bias.}
	\begin{tabular}{lcccccccc} 
	\hline
Start UT Date  & Passband & $q$ (\%) & $u$ (\%) & $\theta$ ($^{\circ}$) & $p$ (\%) & $q_{\rm ISP}$ & $u_{\rm ISP}$& $p_{\rm ISPcorr}$ (\%) \\
\hline
2020-01-03 03:52:16 & $V$ & 3.30  $\pm$ 0.63 & 0.05 $\pm$ 0.67 & 0.4 $\pm$ 5.5 & 3.18 $\pm$ 0.63  & -0.20 $\pm$ 0.07 & 0.51 $\pm$ 0.08 & 3.42 $\pm$ 0.64 \\ 
2020-01-03 05:02:54 & $R$ & 3.67 $\pm$ 0.82 & 0.36 $\pm$ 0.95 & 2.8 $\pm$ 6.4 & 3.51 $\pm$ 0.82 & -0.48 $\pm$ 0.09 & 0.50 $\pm$ 0.12 & 3.99 $\pm$ 0.83\\
2020-01-03 05:32:44 & $I$ & 3.97 $\pm$ 0.48 & 0.28 $\pm$ 0.38 & 2.0 $\pm$ 3.5 & 3.92 $\pm$ 0.48 & -0.59 $\pm$ 0.09 & 0.30 $\pm$ 0.05 & 4.51 $\pm$ 0.49\\ 
\hline
	\end{tabular}\\
\end{table*}


\subsection{Light curve of AT\,2019xis}

\begin{figure}
\centering
\includegraphics[width=\columnwidth]{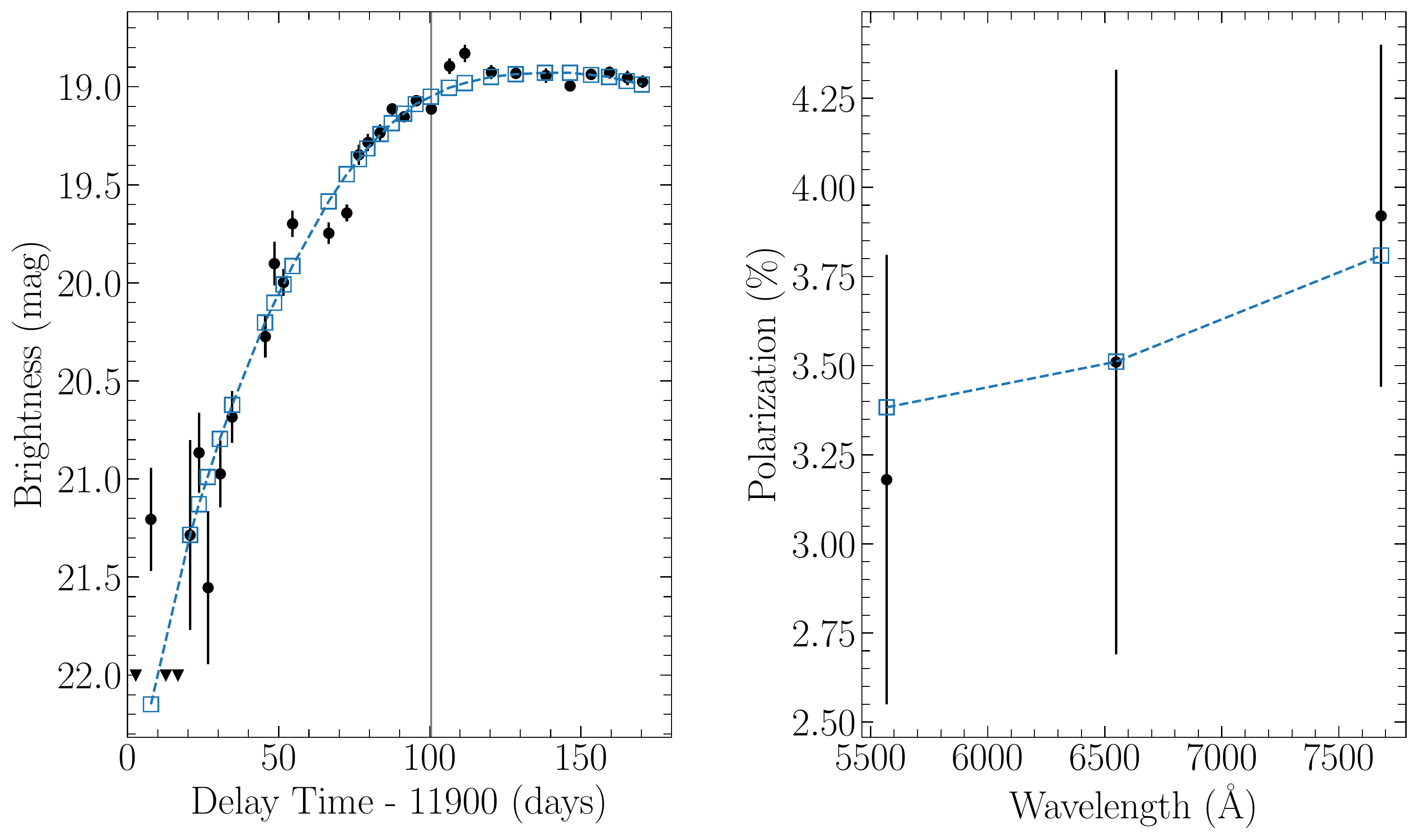}
\caption{Light curve in the $I$ band (left panel) and polarization measurements (right panel) of the light echo AT\,2019xis. Also shown is the simultaneous best fit with the simulated model (blue squares and dashed lines in both panels). The vertical line in the left panel marks the time of the polarization observations, and the triangles are non-detections. The observations shown in this plot were not corrected due to ISP or extinction (see Case i in Sect.~\ref{sect:model}).
}
\label{fig:VRIMag1}
\end{figure}

We used the publicly available light curve\footnote{\href{http://ogle.astrouw.edu.pl/ogle4/transients/transients.html}{http://ogle.astrouw.edu.pl/ogle4/transients/transients.html}} of AT\,2019xis from the Optical Gravitational Lensing Experiment (OGLE) project database \citep{2015AcA....65....1U}. The light echo was observed as part of their OGLE-IV sky survey with the 1.3 m Warsaw University Telescope located at the Las Campanas Observatory in Chile in the $I$-band at 29 epochs between 2019-10-02 and 2020-03-13. There are also upper limit (non-detection) measurements available before 2019-10-02. The light curve is shown in Figure~\ref{fig:VRIMag1}.

\subsection{Interstellar polarization and dust extinction}
\label{sect:isp_ext}

The LMC itself is very dust rich and there is also some Galactic foreground dust along the line of sight toward the LMC. \citet{1990AJ.....99.1483F} found a total reddening along the SN\,1987A sight line of $E(B-V)$ $\simeq$ 0.16 mag, with the LMC component of $E(B-V) \simeq$ 0.10 mag and the Galactic component of $E(B-V) \simeq$ 0.06 mag. This is consistent with the result by \citet{1990PASP..102..131W} who determined an $E(B-V)$ = 0.17 $\pm$ 0.02 mag from photometry of stars near SN\,1987A. Furthermore, \citet{1990AJ.....99.1483F} measured the reddening of the star Sk-69$^\circ$ 203, $E(B-V)$ = 0.19 mag, located in the 30 Doradus region 2.2\arcmin\, north of the SN\,1987A position.

There are also polarization measurements along the line of sight to SN\,1987A available in the literature. \citet{1976A&AS...24..357S} reported the Galactic foreground interstellar polarization (ISP) p = 0.40 $\pm$ 0.13 \% and a position angle of 20\degr\, for the region where SN\,1987A is located. \citet{1991ApJ...375..264J} reported broad-band polarimetry of SN\,1987A taken at the La Plata Observatory by Mendez (priv. comm.) from late 1988 to early 1989. Mendez determined ISP along the sight line to SN\,1987A that can be described with the following Serkowski parameters \citep{serk1975}: p$\rm _{max}$ $\simeq$ 1.05 \%, $\rm \lambda_{max} \simeq$ 5400 \AA\, and K $\simeq$ 1.1 and a polarization angle of $\theta \simeq$ 34$^\circ$.

The dust cloud from which the light echo propagates is, however, located 4.\arcmin05 (about $\sim$200 pc, \citealt{2019TNSAN.163....1T}) East from SN\,1987A . In this region the dust properties along the sight line may differ compared to the sight line toward SN\,1987A. 

To investigate the dust extinction towards AT\,2019xis we examined the publicly available MUSE data of AT\,2019xis (Prog. ID 2104.D-5041(A), PI Kuncarayakti) and looked for the Na I D $\lambda$5890\,\AA\, and $\lambda$5896 \AA\, absorption lines which are well known tracers of dust and gas in the Milky Way \citep{2012MNRAS.426.1465P}. The spectra of AT\,2019xis show no obvious Na I D lines, however a bright (14.4 G mag) star, Gaia EDR3 4657665228220710912, located 3.5\arcsec\, West from AT\,2019xis, displays weak Na I D lines. The median of the geometric distance posterior from the Earth to the star is r$_{\rm med}$geo $\sim$ 14.7 kpc \citep{2021AJ....161..147B}. Therefore, this line of sight is expected to include a large fraction of Galactic dust. 
The equivalent widths (EW) of the Na I D absorption lines are EW(D2) = 0.23 $\pm$ 0.01 \AA\, and EW(D1) = 0.22 $\pm$ 0.01 \AA. Following the empirical relation (valid for Milky Way dust) between the sodium absorption and dust extinction by \citet{2012MNRAS.426.1465P} this corresponds to a reddening of $E(B-V)$ = 0.05$\pm$0.01 mag (see their eq. 9). 
Note that this value is comparable to the Galactic reddening component determined by \citet{1990AJ.....99.1483F}.

We also determined the ISP towards AT\,2019xis following the same method as in \citet[][see their Fig.~2]{2022MNRAS.509.6028C}, by calculating the polarization of field stars. The field stars polarization measurements have been corrected for instrumental polarization, which is negligible in the center of the field but increases up to $\sim$1\% at the borders of the field \citep{Patat2006,2020A&A...634A..70G}.  
The ISP of the field was estimated by taking the weighted mean (with inverse-variance weights) of the field stars' Stokes $q$ and $u$, using 44, 59 and 40 field stars in the $V$, $R$ and $I$ band, respectively. The Stokes $q_{\rm ISP}$ and $u_{\rm ISP}$ are listed in Table~\ref{tab:impol_results}.
Despite a wide range of the field stars polarization values in all bands, ranging from $\sim$0.2 to $\sim$ 2 \%, the weighted means in different bands are relatively consistent. The mean polarization in the $V$, $R$ and $I$ bands are p = 0.53 $\pm$ 0.08 \% with a polarization angle of $\theta$ = 55.9 $\pm$ 4.0 degrees; p = 0.68 $\pm$ 0.10 \% with $\theta$ = 66.8 $\pm$ 4.2 degrees; and p = 0.65 $\pm$ 0.08 \% with $\theta$ = 76.6 $\pm$ 3.6 degrees, respectively. We applied the mean ISP values to correct the polarization measurements of AT\,2019xis (see Table~\ref{tab:impol_results}) and run various simulations, with and without the ISP correction, as explained in the next section. We also acknowledge that the limitation of this method is that most of the stars used to estimate the ISP are nearby compared to the large distance toward the LMC. This means that there may be more dust along the line of sight toward the LMC producing polarization, which is not being taken into account.

\subsection{Geometry of the dust cloud}
\label{sect:geometry}

SN 1987's light echoes in particular seem to come from extended dust sheets, of course with some substructure in it, but are in general very large and roughly perpendicular to the line of sight. Many of these light echoes are relatively unchanged for years, i.e. the arclets do not change significantly in shape and apparent motion \citep[e.g.][]{1988Natur.334..135S, 1995ApJ...451..806X, 2005Natur.438.1132R}. 
However, AT\,2019xis is special because it is one of the brighter SN\,1987A light echoes and it appears as a point source. Figure~\ref{fig:lightechoimage} shows that the light echo coincides with a tip of a dust pillar visible in the HST image, which is rising from a large dust structure in the east (not shown in the cut out). The comparison of the observations taken on January 3rd, 2020 and April 20th, 2020 also shows that AT\,2019xis does not change significantly in the apparent position and shape throughout the period of the available photometry from OGLE. Additional observations taken on April 12th, 2021 with the ESO Faint Object Spectrograph and Camera (EFOSC2) mounted on the New Techonlogy Telescope (NTT) show that the AT\,2019xis faded away as the light echo propagated further and is scattering from another overdensity $\sim$5 arcsec east from AT\,2019xis.

AT\,2019xis was observed with MUSE at two epochs, on 2020-02-19 and 2020-04-20 (Prog. ID 2104.D-5041(A), PI Kuncarayakti). A detailed analysis of the MUSE data is out of the scope of this paper, but we used the extracted light echo spectra to investigate the shape and morphology of the dust cloud producing AT\,2019xis. 
We cleaned the light echo spectra using wavelet decomposition (\citealt{1989wtfm.conf..286H}, see also \citealt{2010ApJ...711..711W,2019MNRAS.490..578C}) and matched them to SN\,1987A spectral templates. The templates have a cadance of 0.5 days and were generated by interpolating original SN\,1987A spectra observed at more than 80 epochs by \citet{1988AJ.....95.1087P,1990AJ.....99.1133P}. To match the spectra we followed the methods used in the Supernova Identification tool \citep[SNID,][]{1979AJ.....84.1511T,2007ApJ...666.1024B}.

\begin{figure}
\centering
\includegraphics[width=\columnwidth]{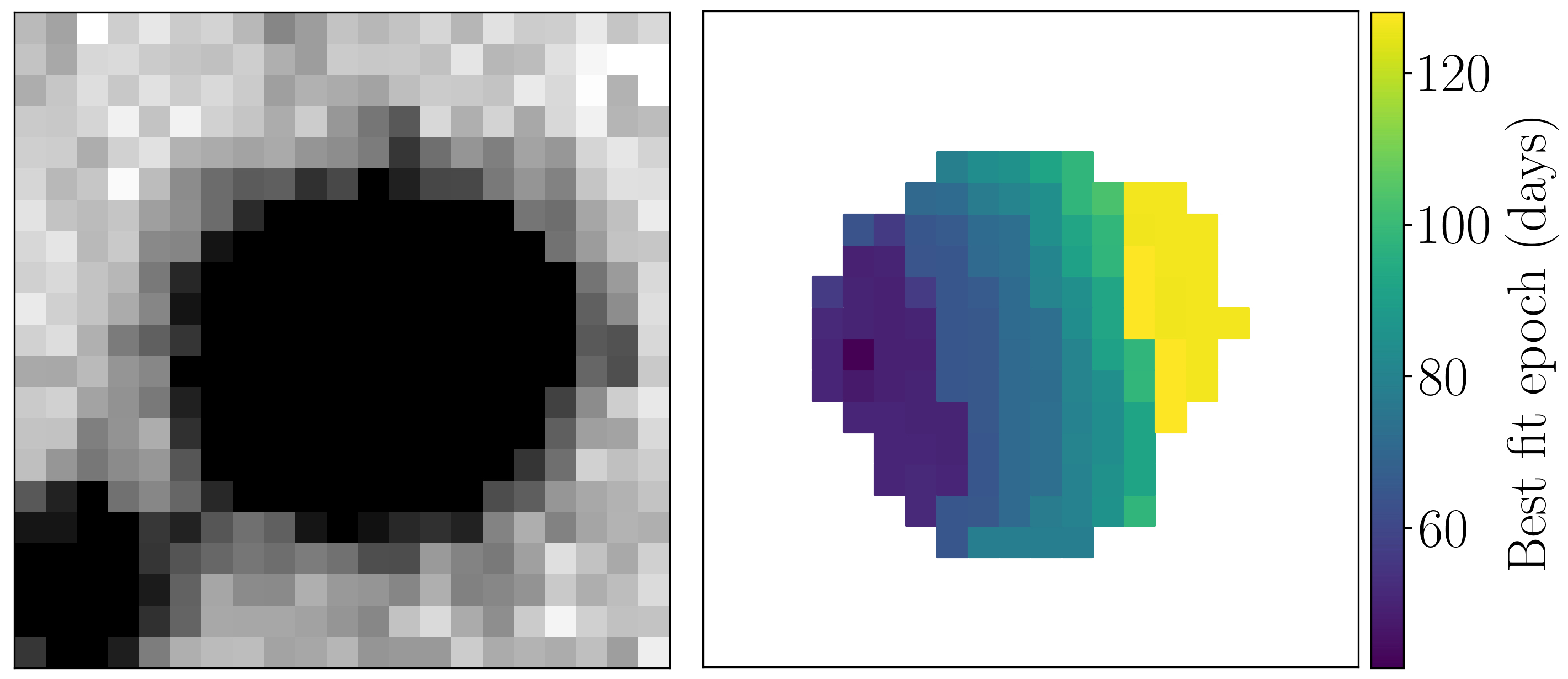}
\includegraphics[width=\columnwidth]{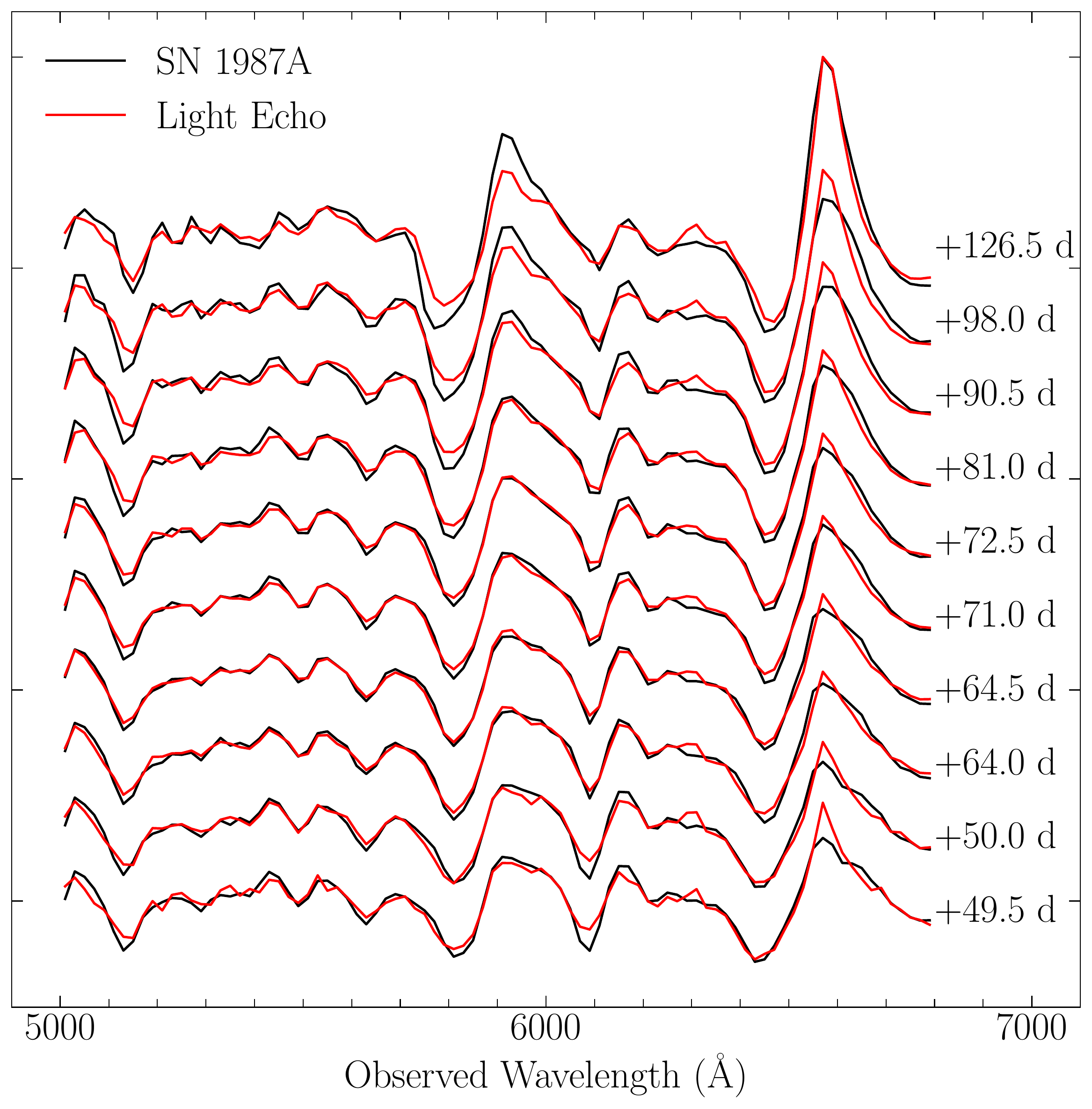}
\caption{\textit{Top left panel:} Light echo AT\,2019xis observed with MUSE on April 20, 2020. The size of the cut out is 4x4 arcsec, and is oriented so that north is up and east is left. We measured the Image Quality FWHM of $\sim$0.7'' ($\sim$3.6 pixel) using neighbouring stars. \textit{Top right panel:} Epoch-map of the light echo determined by matching template spectra of SN\,1987A to each spaxel. The epochs are given relative to the SN explosion. \textit{Bottom panel:} An example sequence of the light echo spectra (red lines) compared to the best matched SN\,1987A templates (black lines). The best fit epochs are indicated.}
\label{fig:MUSEspectra}
\end{figure}

Fig.~\ref{fig:MUSEspectra} shows the epoch-map of the light echo, derived by matching the light echo spectra from each spaxel to SN\,1987A template spectra. The light echo propagates from right to the left in the image (eastwards), and at the diameter of the dust cloud ($\sim$1'') is much smaller than the angular distance to the supernova ($\sim$4.05'), so the light waves are propagating nearly parallel. However, the points of equal epochs display some curvature. The curvature can be explained either due to the spherical shape of the cloud in the case of an optically thick cloud or due to distance differences of the scattering plane in the case of optically thin dust structures. We also note that the light echo spectra match relatively good with the SN\,1987A spectra (see bottom panel in Fig.~\ref{fig:MUSEspectra}), and that the best fit epochs are continuously increasing from east to west, which implies that there is no significant overlapping of the spectra of different epochs, suggesting that the cloud is optically thick. Additionally, the epoch-map is consistent with the dust pillar's spherical appearance visible in the HST image. Thus, assuming a spherical dust cloud seems to be a reasonable simplification for modeling this light echo in the next section.

\section{Model and distance determination}
\label{sect:model}

\citet{Ding2021} used a Monte Carlo radiative transfer model (MCRTM) to simulate light echo observations (light curve and polarization) of SN\,1987A at the position of AT\,2019xis. 
They used $UBVRI$ light curves of SN\,1987A obtained at the Sutherland field station of South African Astronomical Observatory as the input \citep{1987MNRAS.227P..39M, 1987MNRAS.229P..15C, 1988MNRAS.231P..75C, 1989MNRAS.237P..55C, 1988MNRAS.234P...5W} and assumed a simplified dust cloud with similar geometry to AT\,2019xis to study how the simulated observations vary depending on the dust model properties, dust cloud size, and shape. They adopted optical properties of the dust grain models developed by \citet{2001ApJ...548..296W} and investigated the simulated observational differences for different dust grain-size distributions for the Milky Way, LMC and Small Magellanic Cloud (SMC) dust models. 

We adopt the MCRTM by \citet{Ding2021} to determine the distance of the SN\,1987A. 
The apparent brightness and shape of the light curve of the echo is dependent on the intrinsic SN light curve, distance, dust cloud size, shape, optical thickness and the optical properties of the dust grains. Because we know the path length difference between the light of SN\,1987A traveling directly toward Earth and the light echo, the distance of the SN\,1987A determines the scattering angle, with which the light from the supernova is scattered. The scattering geometry was illustrated in \citet{Ding2021}, where the distance to the SN\,1987A was assumed to be 51.4 kpc (Ding, priv. comm.). 
However, assuming that we do not know the distance to SN\,1987A, given the dependence of the light echo's linear polarization on the scattering angle (see also e.g. \citealt{1994ApJ...433...19S}), the scattering angle (and thus, the distance) can be estimated by fitting the MCRTM simulations to the polarization and light curve measurements.

The linear polarization of the light echo is also strongly dependent on the optical thickness \citep{Ding2021}. Optically thick clouds display large spatial variations of the polarization. For example, in an optically thick spherical cloud, the linear polarization increases in the regions closer to the edge of the cloud. 

The dust cloud (AT\,2019xis) distance to the Earth, optical thickness and size of the dust cloud are simultaneously constrained by the light curve and polarization measurements. The distance between the dust cloud and SN 1987A is also constrained by the time delay and the angular distance between AT\,2019xis and SN\,1987A (see \citealt{Ding2021} for details). The dust cloud is assumed to be a homogeneous sphere, whose size and optical thickness are characterized along with the diameter. The dust particles in the cloud are assumed to be prolate spheroids with an aspect ratio of 4. The optical properties of these particles are computed based on the rigorous invariant imbeding T-matrix method \citep{Johnson:88, YangP2019}. A size distribution defined by \citet{2001ApJ...548..296W} as 'LMC' is used to obtain size-distribution averaged dust optical properties. \\

We conduct the MCRTM simulations and fit the observations after applying different extinction and interstellar polarization corrections, as follows: 

(i) Without applying any ISP and extinction correction, i.e. using the original OGLE light curve and the ISP uncorrected polarization measurements. The best fit is shown in Figure~\ref{fig:VRIMag1}.

(ii) same as (i), but assuming halved polarization measurements errors. This test is conducted to investigate the effect of the polarization measurements precision on the distance errors.

(iii) After applying the ISP correction to the polarization measurements determined from field stars (determined in Sect.~\ref{sect:isp_ext}, see also Table~\ref{tab:impol_results}) 

(iv) After applying an ISP correction which follows a Serkowski curve with p$_{\rm max}$=1.05\%, $\rm \lambda_{max}$=5400\AA, and K=1.1, as determined in \citet{1991ApJ...375..264J}.

(v) After applying the ISP correction to the polarization measurements determined from field stars (determined in Sect.~\ref{sect:isp_ext}, see Table~\ref{tab:impol_results}) and a reddening correction to the light curve of $E(B-V)$=0.06 mag assuming Milky Way dust with R$_V$=3.1 (see Sect.~\ref{sect:isp_ext}). Following \citet{1989ApJ...345..245C} this corresponds to the extinction in the $I$-band of A$_I$ = 0.035 mag. For comparison, please note that in the case of LMC dust with R$_V$=2.6, the extinction is very similar, A$_I$ = 0.033 mag.

(vi) After applying the ISP correction to the polarization measurements determined from field stars (see Sect.~\ref{sect:isp_ext}, and Table~\ref{tab:impol_results}) and a reddening correction to the light curve of $E(B-V)$=0.16 mag assuming Milky Way dust with $R_V$=3.1 (see Sect.~\ref{sect:isp_ext}). Following \citet{1989ApJ...345..245C} this corresponds to A$_I$ =0.094 mag. For comparison, assuming LMC dust with R$_V$=2.6, the extinction would be A$_I$ = 0.088.

(vii) - (xi) are same as the case (vi), but instead of using the LMC dust model, we run the fitting for different dust models from \citet{2001ApJ...548..296W}: LMC2, MW3.1, MW4.0, MW5.5 and SMC.

(xii) After applying the ISP correction which follows a Serkowski curve with p$_{\rm max}$=1.05\%, $\rm \lambda_{max}$=5400 \AA, and K=1.1, as determined in \citet{1991ApJ...375..264J}, and a reddening correction to the light curve of $E(B-V)$=0.06 mag assuming Milky Way dust with R$_V$=3.1. Following \citet{1989ApJ...345..245C} this corresponds to A$_I$ = 0.035 mag. 

(xiii) After applying the ISP correction which follows a Serkowski curve with p$_{\rm max}$=1.05\%, $\rm \lambda_{max}$=5400\AA, and K=1.1, as determined in \citet{1991ApJ...375..264J}, and a reddening correction to the light curve of $E(B-V)$=0.16 mag assuming Milky Way dust with $R_V$=3.1 (see Sect.~\ref{sect:isp_ext}. Following \citet{1989ApJ...345..245C} this corresponds to A$_I$ = 0.094 mag.

\section{Results and discussion}
\label{sect:results}

\begin{table*}
	\centering
	\small
	\caption{\label{tab:distance_results} Distance determination results with different models.}
	\begin{tabular}{rlcccccccccc} 
	\hline
  Case  &  Dust & ISP correction &  Extinction correction & Distance & Optical  & Diameter \\
    &  Model  &                 &                        & (kpc)    & thickness$^a$ &  (ly)  \\
\hline
i)  & LMC & No correction & No corrections & 53.44$\pm$2.51 & 3.38$\pm$0.23 & 1.36$\pm$0.05 \\
ii) & LMC & No correction (assuming half errors of p) & No corrections & 53.45$\pm$1.37 & 3.37$\pm$0.23 & 1.36$\pm$0.05  \\
iii) & LMC &  Field stars (this work) & No correction & 49.09$\pm$2.16 & 4.20$\pm$0.44 & 1.29$\pm$0.09 \\
iv) & LMC &  \citet{1991ApJ...375..264J} & No correction  & 58.72$\pm$3.31 & 2.63$\pm$0.23 & 1.50$\pm$0.04 \\
v)  & LMC &  Field stars (this work) & A$_I$= 0.035 mag & 49.30$\pm$2.16 & 3.87$\pm$0.07 & 1.26$\pm$0.01  \\
vi) & LMC & Field stars (this work)& A$_I$= 0.094 mag & 49.67$\pm$2.17 & 3.50$\pm$0.08 & 1.26$\pm$0.02  \\
vii) & LMC2 & Field stars (this work)& A$_I$= 0.094 mag &  49.77 $\pm$2.14   & 3.68 $\pm$0.32     & 1.34 $\pm$0.07 \\
viii)& MW3.1 & Field stars (this work)& A$_I$= 0.094 mag & 54.87 $\pm$2.35 & 1.22 $\pm$0.12   & 1.68 $\pm$0.02 \\
ix)& MW4.0 & Field stars (this work)& A$_I$= 0.094 mag &  52.65 $\pm$2.29  &  2.03 $\pm$0.34    & 1.39 $\pm$0.04 \\
x)& MW5.5 &  Field stars (this work)& A$_I$= 0.094 mag & 49.23 $\pm$2.12 & 3.72 $\pm$0.12   & 1.30 $\pm$0.03  \\
xi) & SMC  &  Field stars (this work)& A$_I$= 0.094 mag & 53.19 $\pm$2.39  &  1.89 $\pm$0.31    & 1.42 $\pm$0.03  \\
xii) & LMC &   \citet{1991ApJ...375..264J} & A$_I$= 0.035 mag & 58.92$\pm$3.34 & 2.46$\pm$0.27 & 1.52$\pm$0.04  \\
xiii) & LMC &  \citet{1991ApJ...375..264J} & A$_I$= 0.094 mag & 59.39$\pm$3.27 & 2.20$\pm$0.28 & 1.57$\pm$0.04  \\
\hline
	\end{tabular}\\
 \textbf{Notes.} $\rm ^a$ The optical thickness is given at 0.8 $\rm \mu m$.
\end{table*}

As shown in the right panel in Fig.~\ref{fig:VRIMag1}, the observed trend in the degree of polarization of scattered light from the dust cloud shows a possible increase toward the red wavelengths. However, we note that the measurement errors are relatively large and that within the errors, the degree of polarization may also be consistent with the rise toward the blue wavelengths. A significant polarization, however, is detected in the $I$ band.

One of the misconceptions of polarization by dust scattering is that polarization is highest in the blue, as quoted in recent papers on the continuum polarization observed along the lines of sight toward Type Ia Supernovae \citep[see e.g.][]{2019MNRAS.490..578C,2017MNRAS.471.2111C,2022MNRAS.509.6028C}. However, this is correct only when the cloud is optically thin at the red wavelengths. 
In contrast, in the case of scattering from optically thick dust clouds, the linear polarization is expected to show a steady increase from 0.3 $\rm \mu m$ to 1.0 $\rm \mu m$ (see \citet{white79, vosh1994, kartje1995, zubko2000}). Such behavior is, for instance, observed in reflection nebulae \citep{zellner1974}.

The observed possible increase in polarization towards the red wavelengths in the case of AT\,2019xis may suggest that the scattering cloud is optically thick in all three bands ($V$, $R$, and $I$). In optically thick clouds, effective scattering only occurs in the outermost layer of the cloud, and the detailed models presented in \citet{Ding2021} demonstrate that polarization can actually peak in the red wavelengths.
Additionally, \citet{Ding2021} has demonstrated that if a scattering cloud with an optical depth close to unity can be resolved spatially, the degree of polarization will exhibit color-dependence at deeper layers of the cloud. Specifically, the degree of polarization in redder filters is higher at deeper layers. Thus, multi-band polarimetry of spatially resolved dust clouds can be utilized as a simple diagnostic tool to determine the optical depth of clouds in general.

\begin{figure}
\centering
\includegraphics[width=\columnwidth]{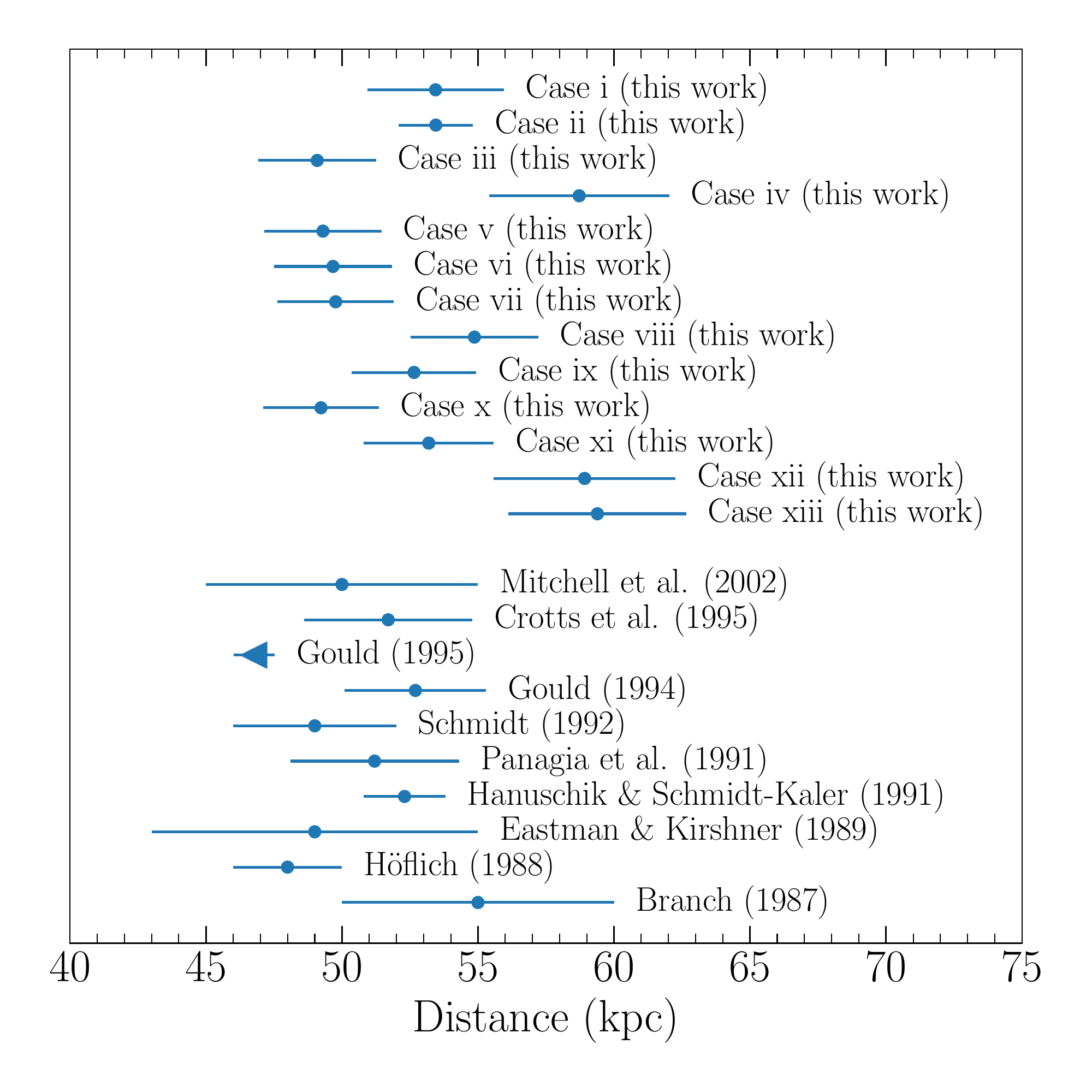}
\caption{Distance measurements to SN\,1987A determined in this work (cases i - xiii), compared to the values from the literature. The distances determined by \citet{1987ApJ...320L..23B}, \citet{1988PASA....7..434H}, \citet{1989ApJ...347..771E}, \citet{1991A&A...249...36H}, \citet{1992ApJ...395..366S} and \citet{2002ApJ...574..293M} are based on the expanding photosphere method, while \citet{1991ApJ...380L..23P}, \citet{1994ApJ...425...51G}, \citet{1995ApJ...452..189G} and \citet{1995ApJ...438..724C} derived the distance based on the circumstellar rings. The distance determined by \citet{1995ApJ...452..189G}, marked with a triangle, is an upper limit. For the description of cases i - xiii, see Sect.~\ref{sect:model} and Table~\ref{tab:distance_results}.}
\label{fig:pol_results_comparison}
\end{figure}

The distances to SN\,1987A were constrained by fitting the MCRTM simulations \citep{Ding2021} to the AT\,2019xis light curve and polarization observations. Examples of the simulated light echo images are shown in Sect. 3.2 in \citealt{Ding2021}.
The results are shown in Figure~\ref{fig:pol_results_comparison} in comparison to literature distances, and summarized in Table~\ref{tab:distance_results}.
Figure~\ref{fig:pol_results_comparison} shows that all the distances are in general comparable to the literature values. Also the distance uncertainties of $\sim$5\% are comparable to the uncertainties achieved with other methods, the expanding photosphere method and distance determinations using the circumstellar rings. However, our uncertainties are highly dependent on the polarization uncertainty.
When comparing the results of cases (i) and (ii), it becomes evident that by reducing the assumed polarization uncertainty by half, the precision of the distance estimate approximately doubles (see Table~\ref{tab:distance_results}).
Furthermore, the degree of polarization also has a high impact on the distance results. After correcting our initial polarization measurements (see case i with no corrections applied) for the ISP as determined by \citet{1991ApJ...375..264J}, we find that the estimates of distances to SN\,1987A increase to $\sim$ 59 kpc (see cases iv, xii, xiii). In contrast, in the cases iii, v, and vi, after correcting our measurements due to the ISP based on the field stars, which effectively leads to an increase of the polarization values (see Table~\ref{tab:impol_results}), the distance estimates to SN\,1987A decreased to $\sim$ 49 kpc.

In contrast, the distance is only minimally affected by the extinction correction of the light curve. This can be seen by comparing case iii, where no extinction correction is applied, to case vi, where an extinction correction of A$_I$=0.094 mag is applied. The resulting distances are 49.09 $\pm$ 2.16 kpc and 49.67 $\pm$ 2.17 kpc without and with the extinction correction, respectively (Table~\ref{tab:distance_results}).

The analysis of cases (vi) - (xi) highlights how different dust models impact distance measurements. When fitting the models with LMC dust, the distance to SN,1987A is estimated to be smaller (around 49 kpc) than when using MW and SMC dust models, which produce distances ranging from 49 to 54 kpc. Although there is some uncertainty introduced by variations in the optical properties of the dust, we consider the use of LMC dust to be a reasonable assumption. This is because the properties of the LMC dust model were derived from independent observations of the LMC, as detailed in \citealt[][]{2001ApJ...548..296W}.
The assumption of the geometry of the scattering dust cloud may also introduce an uncertainty in the distance estimate, but the uncertainty is less than that resulting from the optical properties of the dust particles.

Therefore, assuming the LMC dust model is correct, the ISP correction is the main source of uncertainty in distance estimation. Consequently, the distance to SN\,1987A can be constrained to a range between 49.30 $\pm$ 2.16 kpc (case v) and 59.39 $\pm$ 3.27 kpc (case xiii). The weighted mean distance obtained from cases v, vi xii and xiii is 52.4 $\pm$ 1.3(stat) $\pm$ 4.8(sys) kpc. \\

In summary, we have demonstrated that the use of polarization measurements of light echoes, combined with photometric measurements, can be utilized to estimate distances to supernovae or other transients in our Milky Way and beyond, with a level of accuracy that is comparable to other techniques. However, this method does have some limitations, as it relies on several assumptions regarding dust model properties, dust grain-size distributions, dust cloud size, and the geometric shape of the dust cloud. Despite these limitations, one of the most significant sources of uncertainty is the polarization uncertainty. By reducing this uncertainty, for example, by obtaining more polarimetry observations, it may be possible to achieve greater precision than what can be obtained with other methods.\\

\section*{Acknowledgements}
We would like to thank to Craig J. Wheeler, Ferdinando Patat, Yi Yang, and Heloise Stevance for their contributions and insightful discussions that improved the quality of the paper.
The work of A.C. is supported by NOIRLab, which is managed by the Association of Universities for Research in Astronomy (AURA) under a cooperative agreement with the National Science Foundation. L.W. is partially supported by the NSF through the grant AST 1813825.
This work benefited from L.A.Cosmic \citep{2001PASP..113.1420V}, IRAF \citep{1986SPIE..627..733T}, PyRAF and PyFITS. PyRAF and PyFITS are products of the Space Telescope Science Institute, which is operated by AURA for NASA. We thank the authors for making their tools and services publicly available.  
This work is based on observations collected at the European Organisation for Astronomical Research in the Southern Hemisphere under ESO programs 2104.C-5031(A) and 2104.D-5041(A); and on data products from observations made with ESO Telescopes at the La Silla Paranal Observatory under programme 1106.D-0811(M): PESSTO (the Public ESO Spectroscopic Survey for Transient Objects). The execution in the service mode of these observations by the VLT/NTT operations staff and PESSTO observers is gratefully acknowledged.
Based on observations made with the NASA/ESA Hubble Space Telescope, and obtained from the Hubble Legacy Archive, which is a collaboration between the Space Telescope Science Institute (STScI/NASA), the Space Telescope European Coordinating Facility (ST-ECF/ESA) and the Canadian Astronomy Data Centre (CADC/NRC/CSA).

\bibliography{impolbib}{} 
\bibliographystyle{aasjournal}



\end{document}